# Long and Short Memory in Economics:
# Fractional-Order Difference and Differentiation

**Vasily E. Tarasov**
Skobeltsyn Institute of Nuclear Physics, Lomonosov Moscow State University, Moscow 119991, Russia
E-mail: v.e.tarasov@bk.ru; tarasov@theory.sinp.msu.ru

**Valentina V. Tarasova**
Lomonosov Moscow State Business School, Lomonosov Moscow State University, Moscow 119991, Russia
E-mail: v.v.tarasova@mail.ru

**Abstract**
Long and short memory in economic processes is usually described by the so-called discrete fractional differencing and fractional integration. We prove that the discrete fractional differencing and integration are the Grunwald-Letnikov fractional differences of non-integer order d. Equations of ARIMA(p,d,q) and ARFIMA(p,d,q) models are the fractional-order difference equations with the Grunwald-Letnikov differences of order d. We prove that the long and short memory with power law should be described by the exact fractional-order differences, for which the Fourier transform demonstrates the power law exactly. The fractional differencing and the Grunwald-Letnikov fractional differences cannot give exact results for the long and short memory with power law, since the Fourier transform of these discrete operators satisfy the power law in the neighborhood of zero only. We prove that the economic processes with the continuous time long and short memory, which is characterized by the power law, should be described by the fractional differential equations.

## Introduction: long and short memory processes

Economic processes with long memory are actively studied in recent years (for example, see Teyssiere, G., et. al. (2007), Baillie, R.N. (1996), Banerjee, et. al. (2005), Beran, J. (1994) and Beran, J. et. al. (2013)). Reviews of econometric articles on long memory were suggested by Baillie, R.N. (1996), Robinson, P. M. (2003), Banerjee, et. al. (2005). The mathematical statistics for the long-memory processes has been described in detail by Beran, J. (1994) and Beran, J. et. al. (2013). There are several ways of defining long and short memory of discrete and real-valued time series $y_t$, which are formulated for the time and the frequency domains.

In the time domain, an economic stochastic process $y_t$ exhibits a memory with order $d$, when its autocovariance function (ACF) at lag $k$ satisfies the condition
$$\rho(k) \sim c_\rho \cdot k^{2d-1}, (k \to \infty), \quad (1)$$
where ρ(k) is the autocovariance function (ACF) at lag $k$, and $c_\rho$ is a finite constant. In equation (1) the symbol "~" means that the ratio of the left and right hand sides is finite, when $k$ tends to infinity.

In the frequency domain, an economic stochastic process exhibits a memory with order $d$, when the corresponding spectral density function $S_y(\omega)$ satisfies the condition
$$S_y(\omega) \sim c \cdot \omega^{-2d}, (\omega \to 0), \quad (2)$$
where $S_y(\omega) = |\hat{y}(\omega)|^2$ is the spectral density of the process $y_t$ (time series), and $\hat{y}(\omega) := (Fy_t)(\omega)$ is its Fourier transform. In equation (2) the symbol "~" means that the ratio of the left and right hand sides is finite, when ω tends to zero.

The parameter $d$ is a characteristic of the memory of the economic process, which is described by time series $y_t$. In particular, when $d>0$ the spectral density function (2) is unbounded in the neighborhood of zero and such economic process is called a long memory process. A short-memory process is characterized by power law decaying autocovariance function at $k \to \infty$ and the power law behavior (2) of the spectral density function at $\omega \to 0$, that corresponds to d<0.

There are different methods of memory estimation. One of the most actively used methods is semiparametric estimation, which is based on the spectral density function in the neighborhood of zero according to condition (2). Semiparametric estimators are based on the information included in a periodogram, but for very low frequencies. Note that this restriction of only low frequencies leads to insensitivity of these estimators with respect to different short term shocks.

## Fractional differencing and fractional difference of non-integer order

Long memory was first related to fractional differencing and integrating by Granger, C.W.J., Joyeux, R. (1980), and Hosking, J.R.M. (1981), using the discrete time stochastic process (see also Parke, W.R. (1999), Ghysels, E., et. al. (2001), Gil-Alana, L.A., et. al. (2009)). Granger and Joyeux, and Hosking independently propose the so-called autoregressive fractional integrated moving average model (ARFIMA models). We say that $\{y_t, t = 1, 2, \ldots, T\}$ is an ARFIMA (0, d, 0) model if we have the following equation of discrete time stochastic process
$$(1-L)^d y_t = \varepsilon_t, \quad (3)$$
where L is the lag operator ($Ly_t = y_{t-1}$), $d$ is the order of the fractional differencing (integration), which need not be an integer, $y_t$ is the stochastic process, and $\varepsilon_t$ is independent and identically distributed (i.i.d.) white noise process of random variables with mean $E(\varepsilon_t) = 0$ and variance $V(\varepsilon_t) = \sigma_\varepsilon^2$.

The expression $(1-L)^d$ can be defined by the series expansion, Samko, S.G., et. al. (1993), p. 371, in the form
$$(1-L)^d := \sum_{m=0}^{\infty} (-1)^m \cdot \binom{d}{m} \cdot L^m, \quad (4)$$
where $\binom{d}{m}$ are the generalized binomial coefficients (see equation 1.50 of the book, Samko, S.G., et. al. (1993), p. 14, that are defined by the equation
$$\binom{d}{m} := \frac{\Gamma(d+1)}{\Gamma(d-m+1) \cdot \Gamma(m+1)}. \quad (5)$$

where $\Gamma(z)$ is the Euler gamma function. We can write (4) as
$$(1-L)^d = 1 - d \cdot L - \frac{1}{2} \cdot d \cdot (1-d) \cdot L^2 - \frac{1}{6} \cdot d \cdot (1-d)(2-d) \cdot L^3 - \cdots \qquad (6)$$

Using equation 1.48 of the book, Samko, S.G., et. al. (1993), p. 14, the binomial coefficients $\binom{d}{m}$ can be written in the form
$$\binom{d}{m} := \frac{(-1)^{m-1} \cdot d \cdot \Gamma(m-d)}{\Gamma(1-d) \cdot \Gamma(m+1)}. \qquad (7)$$
Using (7), equation (4) can be represented in the following form
$$(1-L)^d = \sum_{m=0}^{\infty} \frac{\Gamma(m-d)}{\Gamma(-d) \cdot \Gamma(m+1)} \cdot L^m, \qquad (8)$$
which is usually used in the econometric papers on long memory and time series.

It should be noted that the operator
$$\Delta^d := (1-L)^d \qquad (9)$$
is the difference of fractional (integer or non-integer) order, which is called the Grunwald-Letnikov fractional difference of order d with the unit step T=1, Samko, S.G., et. al. (1993), Podlubny, I. (1998), Kilbas, A.A., et. al. (2006).

The Grunwald-Letnikov fractional difference $\Delta_T^\alpha$ of order $\alpha$ with the step T is defined by the equation
$$\Delta_T^\alpha y(t) := (1-L_T)^\alpha y(t) = \sum_{m=0}^{\infty} (-1)^m \cdot \binom{d}{m} \cdot y(t - m \cdot T), \qquad (10)$$
where $L_T y(t) = y(t-T)$ is fixed-time delay and the time-constant T is any given positive value.

As a result, equation (3) is the fractional difference equation. Equation (3) can be generalized for the continuous time case by using the fractional difference equation
$$\Delta_T^\alpha y(t) = \varepsilon(t). \qquad (11)$$
It should be noted that the Grunwald-Letnikov fractional difference (10) may converge for $\alpha<0$, if $y(t)$ has a "good" decrease at infinity, Samko, S.G., et. al. (1993), p. 372. For example, we can use the functions $y(t)$ such that $|y(t)| \leq c \cdot (1+|t|)^{-\mu}$, where $\mu>|\alpha|$. This allows us to use (10) as a discrete fractional integration in the non-periodic case.

As a result, equation (3) of ARFIMA model is the fractional difference equation with the Grunwald-Letnikov fractional difference (10) of order $\alpha=d$.

**Continuously distributed long and short memory and fractional differential equations**

To describe continuously distributed long and short memory, we can apply fractional-order derivatives instead of fractional-order differences. The Grunwald-Letnikov fractional difference (10) allows us to define the fractional-order derivatives, Samko, S.G., et. al. (1993), p. 373, by the equations
$$^{GL}_{\pm}D^\alpha Y(t) := \lim_{h \to +0} \frac{1}{T^\alpha} \Delta^\alpha_{\pm T} Y(t). \qquad (12)$$
The fractional derivatives $^{GL}_{\pm}D^\alpha$ are called the Grunwald-Letnikov fractional derivatives of order $\alpha$, Samko, S.G., et. al. (1993), Podlubny, I. (1998), Kilbas, A.A., et. al. (2006).

Using the fractional-order derivatives (12), equation (11) can be generalized for the continuous time case. Memory processes with continuous time can be described by the equation
$$^{GL}_{+}D^\alpha Y(t) = E(t). \qquad (13)$$
Equation (13) is the fractional differential equation with derivatives of order $\alpha$, Podlubny, I. (1998), Kilbas, A.A., et. al. (2006).

Note that the Gurnwald-Letnikov derivatives (12) coincide with the Marchaud fractional derivatives (Theorems 20.2 and 20.4 of Samko, S.G., et. al. (1993)). The Grunwald-Letnikov and Marchaud derivatives have the same domain of definition. The Marchaud fractional derivatives coincide with the Liouville fractional derivatives for a wide class of functions, Samko, S.G., et. al. (1993), p. 110-111. It is important to emphasize that the Fourier transform of the Liouville fractional integral and derivative has the power law form, Kilbas, A.A., et. al. (2006), p. 90, and, Samko, S.G., et. al. (1993), p. 137. The Fourier transform of the Liouville fractional integral is
$$F\{^L_{\pm}I^\alpha Y(t)\}(\omega) = (\mp i\omega)^{-\alpha} F\{Y(t)\}(\omega). \qquad (14)$$

The Fourier transform of the Liouville fractional derivative is represented by the expression
$$F\{{}_{\pm}^{L}D^{\alpha}Y(t)\}(\omega) = (\mp i\omega)^{\alpha} F\{Y(t)\}(\omega). \tag{15}$$

As a result, we can see that the Fourier transform of the Liouville fractional derivatives and integrals have the power law exactly. We emphasize that the fractional differencing (4), (9) and the fractional-order difference (10) satisfy a power law only asymptotically at $\omega \to 0$, Samko, S.G., et. al. (1993), p. 373.

**Long and short memory with power law and exact fractional differences**

Most of the empirical papers consider the case, when the power law behavior of the spectral density function exists at the zero frequency, i.e. $S_y(\omega) \sim c \cdot \omega^{-2d}$ at $\omega \to 0$. Let us consider an economic process *y(t)* with discrete or continuous time such that its spectral density function $S_y(\omega)$ satisfies the power law exactly for all frequencies. This means that the condition
$$S_y(\omega) = c \cdot \omega^{-2\alpha} \tag{16}$$
holds for all ω›0, where $S_y(\omega) = |\hat{y}(\omega)|^2$ is the spectral density of the process *y(t)* and $\hat{y}(\omega) := (Fy(t))(\omega)$ is the Fourier transform of *y(t)*. In this case, we will call that the process *y(t)* exhibits a memory with power law of the order α. The spectral density function of such economic process is exactly the power function.

It is known that in the non-periodic case the Fourier transform *F* of $\Delta_T^{\alpha} y(t)$ is given, Samko, S.G., et. al. (1993), p. 373, by the formula
$$F\{\Delta_T^{\alpha} y(t)\}(\omega) = (1 - \exp(i\omega T))^{\alpha} F\{y(t)\}(\omega). \tag{17}$$
As a result, the Grunwald-Letnikov fractional difference $\Delta_T^{\alpha}$ of order α cannot correctly describe the processes with of power law memory of order α. The fractional differences (10) cannot be considered as an exact discrete (difference) analog of the Liouville fractional derivative and integrals, Samko, S.G., et. al. (1993), Podlubny, I. (1998), Kilbas, A.A., et. al. (2006), since the Fourier transform of the Grunwald-Letnikov fractional differences are not the power law, i.e.
$$F\{\Delta_T^{\alpha} y(t)\}(\omega) \neq (i\omega T)^{\alpha} F\{y(t)\}(\omega) \tag{18}$$
that leads to the inequality $S_y(\omega) \neq c \cdot \omega^{-2\alpha}$.

In order to have equality in (18), we can use the exact fractional differences that are suggested by Tarasov, V. E. (2016a, 2014, 2015a), and then considered in the papers, Tarasov, V. E. (2015b, 2016b). The kernel $K_{\alpha}(m)$ of the exact fractional differences $\Delta_{T,exact}^{\alpha}$ is expressed by the generalized hypergeometric functions $F_{1,2}(a;b,c;z)$ instead the gamma functions in (5). This kernel of the exact fractional differences is represented by the equation
$$K_{\alpha}(m) := \cos\left(\frac{\pi\alpha}{2}\right) \cdot K_{\alpha}^{+}(m) + \sin\left(\frac{\pi\alpha}{2}\right) \cdot K_{\alpha}^{-}(m), \tag{19}$$
where
$$K_{\alpha}^{+}(m) := \frac{\pi^{\alpha}}{\alpha+1} F_{1,2}\left(\frac{\alpha+1}{2}; \frac{1}{2}, \frac{\alpha+3}{2}; -\frac{\pi^2 m^2}{4}\right), (\alpha > -1), \tag{20}$$
$$K_{\alpha}^{-}(m) := -\frac{\pi^{\alpha+1} \cdot m}{\alpha+2} F_{1,2}\left(\frac{\alpha+2}{2}; \frac{3}{2}, \frac{\alpha+4}{2}; -\frac{\pi^2 m^2}{4}\right), (\alpha > -2). \tag{21}$$
The generalized hypergeometric function $F_{1,2}(a;b,c;z)$ is defined as
$$F_{1,2}(a;b,c;z) := \sum_{k=0}^{\infty} \frac{\Gamma(a+k) \cdot \Gamma(b) \cdot \Gamma(c)}{\Gamma(a) \cdot \Gamma(b+k) \cdot \Gamma(c+k)} \cdot \frac{z^k}{k!}. \tag{22}$$
Using equation (22), the kernel (19) can be represented in the form
$$K_{\alpha}(m) = \sum_{k=0}^{\infty} \frac{(-1)^k \cdot \pi^{2k+\alpha+\frac{1}{2}} \cdot m^{2k}}{2^{2k} \cdot k! \cdot \Gamma\left(k+\frac{1}{2}\right)} \cdot \left(\frac{\cos\left(\frac{\pi\alpha}{2}\right)}{\alpha+2k+1} - \frac{\pi \cdot m \cdot \sin\left(\frac{\pi\alpha}{2}\right)}{(\alpha+2k+2)(2k+1)}\right). \tag{23}$$
The exact fractional difference is defined, Tarasov, V. E. (2016a), by the equation
$$\Delta_{T,exact}^{\alpha} y(t) := \sum_{m=-\infty}^{\infty} K_{\alpha}(m) \cdot y(t - m \cdot T), \tag{24}$$
For α‹0 equation (24) with kernel (19) defines the discrete fractional integration.

Note that Hosking used the hypergeometric functions to describe two-parameter ARIMA*(p,d,q)* processes, Hosking, J. R. M. (1981), p. 172. These processes are most conveniently expressed in terms of the hypergeometric functions, Hosking, J.R.M. (1981).

We should emphasize that the Fourier transform $F$ of the exact fractional differences (24) with kernel (19) has the power law exactly, i.e. the equality

$$F\{\Delta_{T,exact}^{\alpha} y(t)\}(\omega) = (i\omega T)^{\alpha} F\{y(t)\}(\omega) \qquad (25)$$

holds as opposed to the fractional differencing (4), (8) and fractional difference (10), where we have inequality (18). Equation (25) means that the spectral density function $S_y(\omega)$ satisfies the power law (16) exactly.

The exact fractional difference can be considered as an exact discrete analog of the Liouville fractional derivatives and integrals Tarasov, V. E. (2016a), since these operators because these operators have the same power-law behavior (14), (15), and (25). Moreover the exact fractional differences for integer orders and the standard derivatives of integer orders have the same algebraic properties, Tarasov, V. E. (2016a), in contrast to the standard finite-differences integer order, Tarasov, V. E. (2015).

Conclusion

As a result, we can conclude that the long and short memory with power law (16) should be described by the exact fractional-order differences, which demonstrate the power law (25). The fractional differencing (4), (8), which are the Grunwald-Letnikov fractional differences (10), cannot give exact results for the long and short memory with power law (16), since these discrete operators satisfy inequality (18). These discrete operators lead us to insensitivity of these mathematical tools with respect to different short term shocks, since the Fourier transform of these difference operators satisfy power law in the neighborhood of zero only. The correct description of the discrete time long and short memory with power law should be based on the exact fractional differences that are suggested in the papers, Tarasov, V. E. (2016a), Tarasov, V. E. (2016b), see also Tarasov, V. E. (2014), Tarasov, V. E. (2015a),. The continuous time description of the economic processes with the long and short power law memory should be based on the fractional derivatives and integrals, and the fractional differential equations, Samko, S.G., et. al. (1993), Podlubny, I. (1998), Kilbas, A.A., et. al. (2006), This mathematical tool allows us to get new fractional dynamical models, Tarasov, V. E. (2010), for economic processes with memory, Tarasova, V. V., et. al. (2016, 2017).